\documentclass[pre,twocolumn,showpacs,preprintnumbers,amsmath,amssymb]{revtex4}      
\usepackage{epsfig}%
\usepackage{longtable}
\begin{document}
\setlength{\topmargin}{-0.25in}
\preprint{U. of IOWA Preprint}

\title{Small numerators canceling small denominators of the high-temperature scaling variables: a systematic explanation in 
 arbitrary dimensions}


\author{Y. Meurice}
\email[]{yannick-meurice@uiowa.edu}
\affiliation{Department of Physics and Astronomy\\ The University of Iowa\\
Iowa City, Iowa 52242 \\ USA
}


\date{\today}

\begin{abstract}

We describe a method to express the susceptibility and higher derivatives of the free energy in terms of 
the scaling variables
(Wegner's nonlinear scaling 
fields) associated with the high-temperature (HT) fixed point of Dyson hierarchical model 
in  arbitrary dimensions.
We give a closed form solution of the linearized problem.
We check that up to order 7 in the HT expansion, all the poles (``small denominators")
that would naively appear in some positive dimension are 
canceled by zeroes (``small numerators"). 
The requirement of continuity in the dimension can be used to lift 
ambiguities which appear in calculations at fixed dimension.
We show that the existence of a HT phase in the infinite volume limit for a continuous 
set of values of the dimension, requires that this mechanism works to all orders. 
On the other hand, most poles at negative 
values of the dimensional parameter (where the free energy density is not well-defined, but RG flows can be studied) persist
and reflect the fact that for special negative values of the dimension, finite-size corrections become leading terms.
We show that the inverse problem is also free of small denominator problems and that the initial values
of the scaling variables can be expressed in terms of the infinite volume limit of the susceptibility and higher derivatives of the free energy. We discuss the existence of 
an infinite number of conserved quantities (RG invariants) and their relevance for the calculation of universal ratios of critical amplitudes.

\end{abstract}


\maketitle

\section{Introduction}

In many problems, one faces the challenge of deriving the macroscopic consequences of 
a microscopic theory. As we look at the problem at increasingly large scales, a sequence of effective 
theories appear and under some appropriate conditions, an infinite volume limit can be taken.
A general method that allows us to construct these flows in the space of theories is the Renormalization 
Group (RG) method \cite{wilson71b}. The study of some RG fixed points and of the linearized flows close to these fixed points has 
produced a successful picture of the universal behavior in second order phase transitions. 
On the other hand, controlling the RG flows beyond the linearized approximation and calculating the related nonuniversal behavior are more difficult issues. This is unfortunately necessary to calculate the critical 
amplitudes. 

As a first step, one can deal with the nonlinear RG flows for simplified models 
where the RG transformation can be implemented without major technical difficulty.
One possibility is to use approximate versions of the exact RG equations \cite{bagnuls00,berges00} such as 
the local potential approximation \cite{hasenfratzlpa}.
Another possibility to address nonlinear questions \cite{collet78,kw88,jsp02} is to use Dyson's hierarchical model \cite{dyson69,baker72}. 
In the following, we use this lattice model for which the block-spin method can be easily implemented. 
This model is briefly reviewed in section \ref{sec:model}.
Other approaches of nonlinear aspects of the RG flows can be found, for instance, in Refs. \cite{gaite96,pelissetto98,pelissetto00,Caracciolo01}.

In the context of ordinary differential equations, a standard method \cite{arnold88} 
to go beyond the linearized approximation in the vicinity of a fixed point, 
consists in constructing a new system of coordinates where the equations become linear. 
However, this type of procedure is often plagued with the ``small denominator problem" initially encountered 
by Poincar\'e in his study of perturbed integrable Hamiltonians. 
In the context of the RG method, these 
new coordinates are called the scaling variables (or the nonlinear scaling fields) 
and were first introduced by Wegner \cite{wegn}.
Recently, we have proposed an ab-initio calculation of the critical amplitudes in 
the high-temperature (HT) phase 
of this model \cite{jsp02}. In this calculation, the critical amplitudes are 
RG-invariant made out of the nonlinear scaling variables associated with Wilson's 
nontrivial IR fixed point \textit{and} the nonlinear scaling variables associated with 
the HT fixed point. 
In this approach, the two fixed points are in some approximate sense dual 
\cite{dual} to each others. 
The scaling variables associated with Wilson's fixed point have been extensively discussed, but much less is known about those associated with the HT fixed point. We emphasize that being able to use both kind of variables is quite convenient for the study of the RG flows in the intermediate region between the two fixed points.

At first sight, the construction of the scaling variables associated with the 
HT fixed point is impossible for $D=3$ and more generally for rational values of $D$, 
because some of the denominators are exactly zero. However, a numerical study in $D=3$ showed \cite{smalld} that in all of the 36 zero denominators considered, a zero numerator miraculously appears. This strongly indicates the 
existence of a general 
mechanism enabling us to overcome the small denominator problem.

In this article, we show that such a mechanism exists and is closely related to the 
existence of the infinite volume limit of the susceptibility and higher derivatives of the free energy. In addition, we address the issue that whenever zero numerator and denominator appear at the 
same time, the coefficients of the nonlinear expansion appear to be undetermined $(\frac{0}{0})$.
We show that this indeterminacy can be lifted by a procedure similar to the
dimensional regularization \cite{thooft73} used for the evaluation of Feynman diagrams.
It should however be emphasized that it is not used here to take care of a UV problem,  since we will be working with a lattice model. 
We will consider the construction of the HT scaling field in  arbitrary dimensions. In our construction, the 
zero denominators appear as poles at particular dimensions and one can study the mechanism of cancellation close
to a pole but not exactly at the pole. 

For Dyson's hierarchical model, the dimension $D$ appears in a continuous parameter $c=2^{1-2/D}$ introduced 
explicitly in section \ref{sec:model}.
The infinite volume limit is well defined for $0<c<2$, or in other words $D>0$. 
The linear variables associated with the HT fixed point are 
introduced in section \ref{sec:linear}. 
A closed form expression for the linear transformation which diagonalizes the linear 
RG transformation is given in  arbitrary dimensions.
The restriction to the first $l_{max}$ of these variables 
can be interpreted as a HT expansion. In section \ref{sec:hofy}, we expand the linear variables in terms 
of the scaling variables . 
We show that up to order 7 in the HT expansion, the poles corresponding to zero denominators in positive dimensions 
($0<c<2$) are exactly canceled by a zero at the numerator. 
The coefficients of the expansion are then unambiguously defined 
rational functions of $c$ with no poles for $0<c<2$. Their poles appear only at negative values of the dimension where the statistical mechanics model 
does not have a well defined infinite volume limit.

The linear variables are linear combinations of the average values of the total field
$\sum_x \phi_x$.
In section \ref{sec:conn}, we use this fact to reexpress the connected parts of the average values of the total field divided by the volume, or in other words, the susceptibility and the 
higher derivatives of the free energy density, in terms of the scaling variables.
We show that up to order 7, the linear contribution is the \textit{only} leading term 
in the infinite volume limit. In section \ref{sec:infvol}, we explain why this should
happen to all orders. In section \ref{sec:inte} 
we explain why it guarantees the cancellations discussed in section \ref{sec:hofy} to all orders.

Having showed that it is possible to construct a solution of the RG flows in the 
HT phase, we then need to calculate the initial values of the 
scaling variables in terms of the local measure (for instance, a Ising measure or a Landau-Ginzburg measure) used to specify the statistical mechanics model. 
This amounts to inverting the previous expansions.
In section \ref{sec:init}, we construct the scaling 
variables in terms of the linear variables and show that the coefficients are free of poles for $0<c<2$. 
We also show that, up to numerical constants, the initial values of the scaling variables are the infinite 
volume limit of the susceptibility and higher order derivatives of the free energy 
density.
This concludes our construction of a complete solution of the RG flows in the HT phase. To be precise, we have shown that various expansions can
be constructed order by order without encountering any small denominator problems and 
that it is possible to study empirically the convergence of these series.
In section \ref{sec:ht}, we show 
with an example how everything we have done can be used to calculate the
HT expansion at finite volume. We also check explicitly that it yields results 
in agreement with calculations performed using independent methods \cite{meuricejmp95}. 
In section \ref{sec:rginv}, we discuss the existence of 
an infinite number of conserved quantities and their relevance for the calculation of universal ratios of critical amplitudes.

\section{Dyson's hierarchical model}
\label{sec:model}

In this section, we remind some basic facts about Dyson's 
hierarchical model that will be needed in the following.
For more details, the reader may consult 
Refs. \cite{finite,hyper}.
We consider fields located at 
$2^{n_{max}}$ sites labeled with $n_{max}$
indices $x_{n_{max}}, ... , x_1$, each being 0 or 1. 
We divide the $2^{n_{max}}$ sites into
two blocks, each containing $2^{n_{max}-1}$ sites. If $x_{n_{max}}=0$,
the site is in the first box, if $x_{n_{max}} = 1$, the site is in the
second box and so on.
The non-local part of the energy reads
\begin{equation}
H_{nl} =
-{\frac{1}{2}}\sum_{n=1}^{n_{max}}(\frac{c}{4})^n\sum_{x_{n_{max}},...,x_{n+1}} 
(\sum_{x_n,...,x_1}\phi_{(x_{n_{max}},...x_1)})^2
\end{equation}
The partition function for a constant source $J$ (or external magnetic field) reads 
\begin{equation}
Z(J)=\prod_x \int  d \phi_xW(\phi_x)\exp ({-\beta H_{nl}+J\sum_y\phi_y})
\label{eq:fin}
\end{equation}
We call $W(\phi_x)d \phi_x$ the local measure. The most common examples are the Ising 
measure $W(\phi)=\delta(\phi^2-1)$ or the Landau-Ginzburg measure $W(\phi)=\exp (-A\phi^2-B\phi^4)$.
The RG transformation consists in integrating over the fields keeping their sum constant in increasingly 
large boxes. After each integration the fields are rescaled by a factor $\sqrt{c/4}$ in order to keep the form of $H_{nl}$ identical, 
and the RG transformation generates a flow in the space of local measures.

Note that for a constant configuration where all the fields take the same value $\overline{\phi}$, the nonlocal part of the energy takes the value
\begin{equation}
H_{nl}(\overline{\phi})=-2^{n_{max}}{(\overline{\phi})^2\frac{1}{2}}\sum_{n=1}^{n_{max}}(\frac{c}{2})^n \ . 
\label{eq:vol}
\end{equation}
In the infinite volume limit $(n_{max}\rightarrow\infty)$, the sum converges 
only for $|c|<2$. Proofs of the existence of the 
thermodynamical limit  for a Ising measure \cite{miracle,parisi88} 
require that the energy does not 
scale faster than the number of sites. This means $|c|<2$ for the model considered 
here.

In the following, we will make a change of variables in order to get rid of $\beta$ in front of $H_{nl}$ in 
Eq. (\ref{eq:fin}) and reabsorb it in the local measure.
Our main object of study will be the generating function (obtained by Fourier transforming the local measure)
\begin{equation}
R_n(k)=1+a_{n,1}k^2+a_{n,2}k^4+\dots  \ ,
\end{equation}
with
\begin{equation}
a_{n,l}=(- \beta)^l \frac{1}{2l!}(\frac{c}{4})^{ln}< (\sum_{2^n sites} \phi_x ) ^{2l}> \ .
\label{eq:fulla}
\end{equation}
The RG transformation can be summarized in terms of the recursion formula
\begin{equation}
R_{n+1}(k) = C_{n+1} \exp \left[ -\frac{1}{2}
{\frac{\partial ^2}
{\partial k ^2}} \right]\left[R_{n} \left(\frac{\sqrt{c}\ k}{2} \right) \right]^2 \ .
\label{eq:rec}\end{equation}
We fix the normalization constant $C_{n}$ so that $R_{n}(0) = 1$.

It is important to remember that 
in the notation $a_{n,l}$, the first index refers to the number of RG steps 
and the second to the powers of the total field. 
Sometimes, the number of RG steps $n$ will be omitted, sometimes the vector index $l$ will be replaced by boldface notations.
We use the parametrization $c=2^{1-2/D}$ such that a free
massless field scales in the same way as in a
usual $D$-dimensional theory. For reference, Dyson's parametrization \cite{dyson69},
was $c=2^{2-\alpha}$.
The logarithm of $R$
generates the connected zero-momentum Green's functions at finite volume.
We emphasize that in the following, the temperature dependence has been absorbed in the
initial $R_0(k)$.
For instance, in the case of an Ising measure, $R_{0}(k) = \cos(\sqrt{\beta}k)$. 

In the HT phase, polynomial truncations of order $l_{max}$
in $k^2$ provide rapidly converging approximations \cite{kw88,finite}.
The RG flows can be expressed in terms of a quadratic map in a $l_{max}$ dimensional 
space
\begin{equation}
a_{n+1, l} = \frac{u_{n,l}}{u_{n,0}} \ ,
\label{eq:aofu}
\end{equation}
with
\begin{equation}
u_{n,\sigma} = \Gamma_{\sigma}^{ \mu \nu} a_{n,\mu} a_{n,\nu} \ ,
\end{equation}
and
\begin{equation}
\Gamma_{\sigma}^{ \mu \nu}
= (c/4)^{\mu+\nu}\
\frac{(-1/2)^{\mu + \nu - \sigma}(2(\mu+\nu))!}{
(\mu+\nu-\sigma)!(2 \sigma)!}  \ ,
\label{eq:struct}
\end{equation}
for $\mu+\nu \geq\sigma$ and zero otherwise.
We use ``relativistic'' notations.
The greek indices $\mu$ and $\nu$
go from $0$ to $l_{max}$, while latin indices $i$, $j$ go from 1 to
$l_{max}$. Repeated indices mean summation unless specified differently. 
With the normalization of Eq. (\ref{eq:aofu}), 
$a_n,0=1$ for any $n$ and is not a dynamical variable. Note that a truncation to order 
$l_{max}$ 
is always implicit in the following. However, for reasons that will be explained 
in the coming section, there is no explicit dependence in $l_{max}$.
 
\section{The linear RG transformation}
\label{sec:linear}

In this section we discuss the linearized RG transformation near the HT fixed point $a_i=0$ for all $i\leq 1$.
For small departure from the HT fixed point $\delta a_{n,i}$ the linear RG transformation reads 
\begin{equation}
\delta a_{n+1,i} \simeq {\mathcal M}_i^j\delta a_{n,j}\ ,
\end{equation}
with 
\begin{equation}
{\mathcal M}_i^j =2\Gamma^{j0}_i=2(\frac{c}{4})^j(-\frac{1}{2})^{j-i}\frac{(2j)!}{(2i)!(j-i)!}\ ,
\label{eq:lin}
\end{equation}
for $i\leq j$ and zero otherwise.

The diagonalization of ${\mathcal M}$ is not too difficult because of its upper triangular form.
The spectrum is given by the diagonal elements:
\begin{equation}
\lambda_{(r)}=2(c/4)^{r} \ ,
\label{eq:hteigenv}
\end{equation}
in agreement with Ref. \cite{collet78}.
We need to construct ${\mathcal R}$, a matrix of right eigenvectors, such that
\begin{equation}
{\mathcal M}_{l}^{i} {\mathcal R}^r_i = \lambda_{(r)} {\mathcal R}^r_l \ ,
\label{eq:rdef}
\end{equation}
(with no summation over $r$). For convenience, the columns of ${\mathcal R}$ are ordered 
as the eigenvalues, $0<c<4$ being  assumed.
We will then introduce the linear coordinates $h_{n,l}$ defined by
\begin{equation}
a_{n,l} = {\mathcal R}^r_l h_{n,r}\ ,
\label{eq:aofh}
\end{equation}
and which transform as 
\begin{equation}
h_{n+1,r}\simeq \lambda_{(r)}h_{n,r}
\label{eq:applin}
\end{equation}
in the linear approximation.
The matrix ${\mathcal R}^r_i$ and its inverse are also upper triangular.
This implies that $h_{n,l}$ is of order $\beta ^l$, just as $a_{n,l}$ is.
We will fix the normalization of the 
right eigenvectors in ${\mathcal R}$ in such way that all the diagonal elements are 1.
This guarantees that
$h_{n,l}=a_{n,l}+ {\mathcal O}(\beta ^{l+1})$.

Before entering into the technical details of the construction of ${\mathcal R}$, an 
important consequence of the upper triangular form of ${\mathcal M}$ should be 
noticed. The eigenvectors and eigenvalues of ${\mathcal M}$ are independent of 
a possible truncation. In other words, the fact that ${\mathcal R}$ is upper triangular
means that the polynomial truncations of $R$ to order $k^{2l_{max}}$mentioned in section \ref{sec:model} are indeed
a projection in the subspace spun by the first $l_{max}$ eigenvectors of ${\mathcal M}$. 

We now construct ${\mathcal R}$.
We first notice that for $j>i$,
\begin{equation}
{\mathcal R}_i^j=(\frac{c}{4-c})^{j-i}{\mathcal P}_i^j\ ,
\label{eq:genr}
\end{equation}
with ${\mathcal P}_i^j$  $c$-independent
For indices no larger than 7, the entries of ${\mathcal P}$ are
\[\left(
\begin{array}{ccccccc}
1 & 6 & 45 & 420 & 4725 & 62370 & 945945 \cr 0 & 1 & 
15 & 210 & 3150 & 51975 & 945945 \cr 0 & 0 & 1 & 28 & 630 & 13860 & 3153
15 \cr 0 & 0 & 0 & 1 & 45 & 1485 & 45045 \cr 0 & 0 & 0 & 0 & 1 & 
66 & 3003 \cr 0 & 0 & 0 & 0 & 0 & 1 & 
91 \cr 0 & 0 & 0 & 0 & 0 & 0 & 1 
\end{array}
\right)\]
The matrix ${\mathcal P}$ has remarkable properties 
\begin{subequations}
\label{eq:pid}
\begin{eqnarray}
{\mathcal P}_i^{i+1}{\mathcal P}_{i+1}^{i+2}&=&2{\mathcal P}_{i}^{i+2}
\label{equationa}\\
{\mathcal P}_i^{i+1}{\mathcal P}_{i+1}^{i+3}&+&{\mathcal P}_{i}^{i+2}{\mathcal P}_{i+2}^{i+3}=
{\mathcal P}_i^{i+1}{\mathcal P}_{i+1}^{i+2}{\mathcal P}_{i+2}^{i+3}\label{equationb}
\end{eqnarray}
\end{subequations}
and higher order ones that as we will see are related to the very simple form of
the inverse matrix.
We can condense these relations into the more compact recursion
\begin{equation}
{\mathcal P}_i^{i+m}{\mathcal P}_{i+m}^{i+m+q}=
\left( \begin{array}{c}m+q \cr
q \end{array} \right)
{\mathcal P}_{i}^{i+m+q}\ .
\end{equation}
This implies the closed form expression
\begin{equation}
{\mathcal P}_i^j =(-\frac{1}{2})^{j-i}\frac{(2j)!}{(2i)!(j-i)!}\ .
\label{eq:pexp}
\end{equation}

Using Eq. (\ref{eq:genr}) and (\ref{eq:pexp}), it is easy to check that, as a consequence
of the binomial formula, we have provided  an exact solution of Eq. (\ref{eq:rdef}) with 
the required normalization (ones on the diagonal).
Similarly, one can show that the inverse has the very simple form
\begin{equation}
({\mathcal R}^{-1})_i^j=(-1)^{j-i}{\mathcal R}_i^j \ .
\end{equation}
This equation does not hold for an arbitrary upper triangular matrix. It implies the
identities (\ref{eq:pid}) and many others.

\section{Expression of the linear variables in terms of the scaling variables}
\label{sec:hofy}

In this section, we express the linear variables $h_l$ 
in terms of the (non-linear) scaling variables $y_l$ for which the 
approximate multiplicative transformation of Eq. (\ref{eq:applin}) becomes exact.
If we use ${\mathrm ln}(y_l)$ as our new coordinates, the RG flows become parallel
straight lines. All the dynamics is then contained in the mapping that we now proceed 
to construct.

We first rewrite the RG transformation in the $h_l$ coordinates. Starting with 
the basic Eq. (\ref{eq:aofu}), we replace $a_0$ by 1 and $a_l$ 
by ${\mathcal R}_l^p h_p$. This yields 
\begin{equation}
h_{n+1,l} = \frac{\lambda_{(l)} h_{n,l}
 + \Delta_{l}^{ p q} h_{n,p} h_{n,q} }
{1 + 2\Delta^{p0}_0 h_{n,p}
+ \Delta_{0}^{ p q} h_{n,p} h_{n,q}}\ ,
\label{eq:hrules}
\end{equation}
with coefficients calculable from Eq. (\ref{eq:struct}). For instance,
\[\Delta_{l}^{ p q} =({\mathcal R}^{-1})_l^{l'}\Gamma_{l'}^{p'q'}{\mathcal R}_{p'}^p
{\mathcal R}_{q'}^q \ .\]In general, upper roman indices transform with 
${\mathcal R}$ and the lower ones with $({\mathcal R})^{-1}$.
By construction, the linear transformation is diagonal.

We then introduce the expansion
\begin{equation}
h_{l} = y_l+\sum_{i_{1},i_{2},\ldots} s_{l,i_{1} i_{2} \ldots} y_{1}^{i_{1}}
y_{2}^{i_{2}} \ldots \ ,
\label{eq:nlexph}
\end{equation}
where the sums over the $i$'s run from $0$ to infinity in each variable
with at least two non-zero indices.
In the following, we use the notation ${\mathbf{i}}$ for $(i_1,i_2,\dots)$. 
More generally, vectors will be represented by boldface characters.
The unknown coefficients $s_{l,{\mathbf{i}}}$ in Eq. (\ref{eq:nlexph}) are obtained 
by matching two expressions of $h_{n+1,l}$, one obtained from the RG transformation of the $h_l$ given in Eq. (\ref{eq:hrules}), the other obtained by evolving the scaling variables according to the exact multiplicative transformation
\begin{equation}
y_{n+1,l}=\lambda _{(l)}y_{n,l} \ .
\end{equation}
The matching conditions can be expressed as: 
\begin{equation}
h_{n+1,l}({\mathbf{h}}_n({\mathbf y}))=h_{n,l}(\lambda_1 y_{n,1},\lambda_2 y_{n,2},
\dots ) \ .
\label{eq:match}
\end{equation}
and yield the conditions
\begin{equation}
s_{l, \mathbf{i}} = \frac{N_{l,\mathbf{i}}}{D_{l,{\mathbf i}}}
\ .
\label{eq:denom}
\end{equation}
with
\begin{eqnarray}
\label{eq:num}
N_{l, \mathbf{i}} =& \sum_{\mathbf{j}+\mathbf{k} = \mathbf{i}}
 ( -\Delta_{l}^{ p q} s_{p,\mathbf{j}}
 s_{q,\mathbf{k}} +s_{l,\mathbf{j}} \prod_{m} \lambda_{(m)}^{j_{m}}
2\Delta^{p0}_0 s_{p, \mathbf{k}} )\nonumber \\
&+\sum_{\mathbf{j}
+\mathbf{k}+\mathbf{r}=\mathbf{i}}
 s_{l, \mathbf{j}} \prod_{m} \lambda_{(m)}^{j_{m}} \Delta_{0}^{ p q}\ ,
 s_{p, \mathbf{k}} s_{q, \mathbf{r}} \ .
\end{eqnarray}
and
\begin{equation}
D_{l,{\bf i}}=\lambda_{(l)}-\prod_{m} \lambda_{(m)}^{i_{m}} \ .
\label{eq:deno}
\end{equation}
For a given set of indices $\mathbf{i}$, we
introduce the notation
\begin{equation}
{\mathcal{I}}_q ({\mathbf i})=\sum_m i_m m^q \ .
\label{eq:indices}
\end{equation}
One sees that ${\mathcal{I}}_0$ is the degree of the associated 
product of scaling variables and ${\mathcal{I}}_1$
its order in the HT expansion (since $y_l$ is also of order $\beta ^l$).
Given that all the indices are positive and that at least one index
is not zero,
one can see that if ${\mathbf{j}}+{\mathbf{k}}={\mathbf{i}}$ then
${\mathcal I}_q({\mathbf{j}})<{\mathcal I}_q({\mathbf{i}})$ and
${\mathcal I}_q({\mathbf{k}})<{\mathcal I}_q({\mathbf{i}})$. Consequently,
Eq. (\ref{eq:num}) yields a solution order by order in ${\mathcal{I}}_0$ or
in ${\mathcal{I}}_1$ (since the r.h.s is always contains 
$s_{l,{\mathbf{i}}}$ of lower order in ${\mathcal{I}}_0$ or ${\mathcal{I}}_1$) 
provided that none of the denominators $D_{l,{\bf i}}$ are exactly zero.
The main goal of this article is to investigate what happens when some of the 
denominators happen to be exactly zero.

Using the explicit expression of the eigenvalues Eq. (\ref{eq:hteigenv}), we can 
rewrite the denominators as
\begin{equation}
D_{l,{\bf i}}=2(\frac{c}{4})^l-2^{{\mathcal I}_0({\mathbf i})}
(\frac{c}{4})^{{\mathcal I}_1({\mathbf i})}\ .
\end{equation}
Using the parametrization $c=2^{1-2/D}$,
the zero denominators appear when
\begin{equation}
D-l(D+2)=D{\cal{I}}_0({\mathbf i}) -(D+2){\cal{I}}_1 ({\mathbf i})\ .
\end{equation}
Given that the ${\cal{I}}_q$ are integers, this can only occur at some rational 
values of $D$. Ignoring temporarily this set of values, we can say that for generic values 
of $c$, the denominator is not zero. 
Following the basic idea of dimensional regularization, 
we will then perform, order by order in ${\mathcal I}_1$, the 
construction of the $s_{l,{\mathbf i}}$ for 
a generic value of $c$ and discuss the limit where $c$ takes some special value 
at the end of the calculation.

We now determine the range of values of ${\mathcal{I}}_0$ and ${\mathcal{I}}_1$ relevant
for our problem. 
In Eq. (\ref{eq:nlexph}), we have assumed that $h_l\simeq y_l$ for sufficiently small values 
of the scaling variables. The linear problem is completely solved and we may 
assume ${\mathcal{I}}_0({\mathbf i}) >1$. In addition, since both $h_l$ and $y_l$ are of order $\beta^l$, 
we need ${\mathcal{I}}_1({\mathbf i}) \geq l$. At lowest non-trivial order in $\beta$, we 
have ${\mathcal{I}}_l({\mathbf i})= l$, and 
\[D_{l,{\bf i}}=(\frac{c}{4})^l(2-2^{{\mathcal I}_0({\mathbf i})}) \ . \]
In this special case, the only possible poles are at $c=0$. However, the factor $(\frac{c}{4})^l$ at the denominator is exactly canceled by the same factor appearing 
in the $\Delta_l^{pq}$ in Eq. (\ref{eq:hrules}). More precisely, 
\begin{equation}
h_{n+1,l}=(\frac{c}{4})^l(2h_{n,l}+\sum_{p+q=l}h_{n,p}h_{n,q})+{\mathcal O}(\beta ^{l+1}) \ .
\end{equation}
Using this, it is not difficult to prove by induction that if ${\mathcal{I}}_l({\mathbf i})= l$, 
\begin{equation}
s_{l,\bf i}={\prod_{m} \frac{1}{i_{m}!}} \ .
\label{eq:lowestnt}
\end{equation}
It is thus clear that at the lowest non-trivial order, the coefficients have no singularities.

We now discuss the case ${\mathcal I}_1({\mathbf i})>l$. We have in general
\begin{equation}
D_{l,{\mathbf i}}=2(\frac{c}{4})^l(c_{crit.})^{l-{\mathcal I}_1({\mathbf i})}\  T_{l,{\mathbf i}} ,
\end{equation}
with
\begin{equation}
T_{l,{\mathbf i}}=(c_{crit.}^{{\mathcal I}_1({\mathbf i})-l}-c^{{\mathcal I}_1({\mathbf i})-l})\ ,
\label{eq:tli}
\end{equation}
and
\begin{equation}
c_{crit.}=4\times 2^{(1-{\mathcal I}_0({\mathbf i}))/
({\mathcal I}_1({\mathbf i})-l)}\ .
\end{equation}
One should always keep in mind that $c_{crit.}$ is a function of both $l$ and 
${\mathbf i}$.
Inspection of Eqs. (\ref{eq:struct}) 
and (\ref{eq:genr}) shows 
that the numerator has a factor ${c^{{\mathcal I}_1({\mathbf i})}}{({c-4})^{l-{\mathcal I}_1({\mathbf i})}}$. Consequently,
\begin{equation}
s_{l,\mathbf i}=(\frac{c}{c-4})^{{\mathcal I}_1({\mathbf i})-l}Q_{l,\mathbf i}(c)\ ,
\label{eq:ql}
\end{equation}
where $Q_{l,\mathbf i}(c)$ is a rational function of $c$ 
with no poles or zeroes at 0 or 4. We do not have a compact formula for these
rational functions, however it is easy to calculate them using symbolic 
manipulation programs.

\begin{table*}
\caption{\label{table:ql}Values of $Q_{l,\mathbf i}(c)$, $c_{crit.}$ and $T_{l,\mathbf i}(c)$ defined in the text.}
\begin{ruledtabular}
\begin{tabular}{|c|c|c|c|c|}
\hline
$l$ & $ \prod _m y_m^{i_m}$ & $Q_{l,\mathbf i}(c)$ & $c_{crit.}$ & $-T_{l,\mathbf i}(c)$  \\
\hline
\rule{0mm}{1mm}& & & \\
1&${{y_1}}^2 $  &  $ {\displaystyle \frac{2 + c}{-2 + c} }$  &  $ 2 $  &  $ -2 + c $  \\ 
 \rule{0mm}{1mm}& & & \\
1&${{y_1}}^3 $  &  $ {\displaystyle \frac{-\left( 4 - 20\,c + c^2 \right) }
   {2\,{\left( -2 + c \right) }^2} }$  &  $ 2 $  &  $ -4 + c^2 $  \\
\rule{0mm}{1mm}& & & \\
 1&  $ {y_1}\,{y_2} $  &  $ 
   {\displaystyle \frac{-3\,\left( -40 + c^2 \right) }{-8 + c^2}}$  &  $2\,{\sqrt{2}}$  &  $-8 + 
   c^2 $  \\ 
\rule{0mm}{1mm}& & & \\ 
1& ${{y_1}}^4$  &  ${\displaystyle \frac{-120 + 156\,c - 18\,c^2 + c^3}
   {2\,{\left( -2 + c \right) }^3}}$  &  $2$  &  $-8 + c^3 $  
\\ \rule{0mm}{1mm}& & & \\ 
1& $ {{y_1}}^2\,
   {y_2}$  &  ${\displaystyle \frac{3\,\left( -11520 - 640\,c + 1184\,c^2 + 288\,c^3 + 
       40\,c^4 - 26\,c^5 + 3\,c^6 \right) }{\left( -2 + c \right) \,
     \left( -8 + c^2 \right) \,\left( -16 + c^3 \right) }}$  &  $2\,
   2^{\frac{1}{3}}$  &  $-16 + c^3 $  \\  
\rule{0mm}{1mm}& & & \\ 
1&$ {{y_2}}^2$  &  ${\displaystyle \frac{1536}{-32 + c^3}}$  &  $2\,
   2^{\frac{2}{3}}$  &  $-32 + c^3 $  \\ 
\rule{0mm}{1mm}& & & \\  
1&$ {y_1}\,{y_3}$  &  $15$  &  $2\,
   2^{\frac{2}{3}}$  &  $-32 + c^3 $  \\  
\rule{0mm}{1mm}& & & \\
\hline
\rule{0mm}{1mm}& & & \\
2&$ {{y_1}}^3$  &  ${\displaystyle \frac{6 + c}{2\,\left( -2 + c \right) }}$  &  $1$  &  $-1 + 
   c $  \\ 
\rule{0mm}{1mm}& & & \\ 
2&$ {y_1}\,{y_2}$  &  ${\displaystyle \frac{14 + c}{-2 + c}}$  &  $2$  &  $-2 + c $  \\ 
\rule{0mm}{1mm}& & & \\
2&  $ 
 {{y_1}}^4$  &  ${\displaystyle \frac{-\left( -44 - 28\,c + c^2 \right) }
   {4\,{\left( -2 + c \right) }^2}}$  &  ${\sqrt{2}}$  &  $-2 + c^2 $  \\ 
\rule{0mm}{1mm}& & & \\  
2&$ {{y_1}}^2\,
   {y_2}$  &  ${\displaystyle \frac{-2\,\left( 256 + 304\,c - 112\,c^2 - 14\,c^3 + c^4 \right) }
   {{\left( -2 + c \right) }^2\,\left( -8 + c^2 \right) }}$  &  $2$  &  $-4 + 
   c^2 $  \\ 
\rule{0mm}{1mm}& & & \\  
2&$ {{y_2}}^2$  &  ${\displaystyle \frac{-3\,\left( -104 + c^2 \right) }
   {-8 + c^2}}$  &  $2\,{\sqrt{2}}$  &  $-8 + c^2 $ \\
\rule{0mm}{1mm}& & & \\  
2& $ {y_1}\,{y_3}$  &  ${\displaystyle \frac{240}
   {-8 + c^2}}$  &  $2\,{\sqrt{2}}$  &  $-8 + c^2 $  \\ 
\rule{0mm}{1mm}& & & \\
\hline
\rule{0mm}{1mm}& & & \\
3 &${{y_1}}^4$  &  ${\displaystyle \frac{10 + c}{6\,\left( -2 + c \right) }}$  &  $\frac{1}
   {2}$  &  $-\left(  \frac{1}{2} \right)  + c $  \\ 
\rule{0mm}{1mm}& & & \\  
3&$ {{y_1}}^2\,{y_2}$  &  ${\displaystyle \frac{18 + c}{-2 + c}}$  
   &  $1$  &  $-1 + c $  \\ 
\rule{0mm}{1mm}& & & \\ 
3& $ {{y_2}}^2$  &  ${\displaystyle \frac{16}{-2 + c}}$  &  $2$  &  $-2 + 
   c $  \\ 
\rule{0mm}{1mm}& & & \\ 
3& $ {y_1}\,{y_3}$  &  ${\displaystyle \frac{22 + c}{-2 + c}}$  &  $2$  &  $-2 + c $  \\ 
\rule{0mm}{1mm}& & & \\
\end{tabular}
\end{ruledtabular}
\end{table*}  

Naively, we would expect that 
$Q_{l,\mathbf i}(c)$ has a factor $T_{l,\mathbf i}(c)$ at the denominator and 
other poles inherited from the $s_{l,\mathbf i}$ of lower orders. 
The values of $Q_{l,\mathbf i}(c)$ up to order $\beta ^4$ are 
shown in Table \ref{table:ql}. The naive expectations concerning the poles 
are only observed in 9 cases out of the 17 considered. In the 8 other cases, 
some cancellations occur. For instance, there is no $(c+2)$ at the denominator
of $Q_{1,(3,0,\dots)}$. More importantly, whenever $c_{crit.}<2$, we observe 
a cancellation of {\textit all} the factors appearing in $T_{l,\mathbf i}(c)$. This occurs, for instance, for $Q_{2,(3,0,\dots)}$, where the factors $c-1$ cancel. If we do the calculations explicitly using Eq. (\ref{eq:num}), we 
obtain five terms at the numerator:
\begin{eqnarray}
 N_{2,(3,0,\dots )}=\frac{-11\,c^3}{64}& -& \frac{3\,c^3}{8\,\left( -4 + c \right) }\nonumber \\
  + 
  \frac{c^3}{4\,\left( -4 + c \right) \,\left( -2 + c \right) }&+& 
  \frac{15\,c^4}{64\,\left( -4 + c \right) }\nonumber \\ + 
  \frac{c^4}{8\,\left( -4 + c \right) \,\left( -2 + c \right) }\ \nonumber ,
\end{eqnarray}
while the denominator reads:
\[ D_{2,(3,0,\dots )}=\frac{c^2}{8}(c-1)\]
After reduction and factorization, the numerator becomes:
\[ N_{2,(3,0,\dots )}=\frac{\left( -1 + c \right) \,c^3\,\left( 6 + c \right) }
  {16\,\left( -4 + c \right) \,\left( -2 + c \right) }\ , \]  
canceling the $c-1$ at the denominator. We have pursued the same procedure up
to order  $\beta ^7$ and considered the 175 possible terms. In 50 cases, we had $c_{crit.}<2$.
In each of these 50 cases, we observed  a complete cancellation of $T_{l,{\mathbf i}}(c)$. It seems thus reasonable to conjecture that 
$Q_{l,{\mathbf i}}(c)$ has no poles for $|c|<2$.
If this conjecture is correct,
dimensional regularization provides a unique continuous expression for the coefficients for any $c$ with $|c|<2$ and 
the model is ``solvable" using the recursion for the coefficients given 
by Eq. (\ref{eq:num}). Note that for values of $c$ real and positive, the 
correspondence  
$c=2^{1-2/D}$ implies that the interval $0<c<2$ corresponds to $0<D<+\infty$.
The conjecture implies that for any value of $c$ in this interval, 
we can construct analytical expression of $a_{n,l}$ (which contains
all the thermodynamical quantities)
in terms
of $a_{0,l}$ (which depends on the initial energy density):
\begin{equation}
a_{n,l}=({\mathcal R}^{-1})_l^rh_r(\lambda_1^ny_1({\mathbf a}_0),\lambda_2^ny_2({\mathbf a}_0),
\dots)\ .
\label{eq:solve}
\end{equation}
The initial values of ${\mathbf y}({\mathbf a}_0)$ have a simple 
interpretation discussed in section \ref{sec:init}.
\section{The connected parts}
\label{sec:conn}
The 
generating function of the connected parts of the average values of the total field reads
\begin{equation}
{\rm ln}(R_n(k))=a^c_{n,1}k^2+a^c_{n,2}k^4+\dots \ , 
\label{eq:conngen}
\end{equation}
with 
\begin{equation}
a^c_{n,l}=\sum_{{\mathbf i}:{\mathcal I}_1({\bf i})=l}(-1)^{{\mathcal I}_0({\bf i})-1}({\mathcal I}_0({\bf i})-1)!\prod_m \frac{a_m^{i_m}}{i_m!}\ .
\label{eq:conna}
\end{equation}
We repeat that we are working exclusively in the HT phase and that we do not need to subtract powers of 
the magnetization.
Using Eq. (\ref{eq:aofh}) and the construction discussed in the previous section, we can then calculate $a_{n,l}^c({\mathbf y_n})$. In addition, we have 
\begin{equation}
a_{n,l}^c=(- \beta)^l \frac{1}{2l!}(\frac{c}{4})^{ln}\langle (\sum_{2^n sites} \phi_x) ^{2l}\rangle ^c \ ,
\label{eq:acinter}
\end{equation}
with the connected part of the average values $\langle \rangle ^c$ defined in the usual way. 
For instance,
\begin{eqnarray}
a_{n,2}^c&=&a_{n,2}-(1/2)a_{n,1}^2\nonumber \\
&=&(- \beta)^2\frac{1}{4!}(\frac{c}{4})^{2n}\langle ( \sum_{2^n sites} \phi_x ) ^{4}\rangle^c \nonumber
\end{eqnarray}
with
\begin{eqnarray}
\langle ( \sum_{2^n sites} \phi_x ) ^{4} \rangle ^c&=& \\
\langle ( \sum_{2^n sites} \phi_x ) ^{4} \rangle &-&3 \langle         
(\sum_{2^n sites} \phi_x ) ^{2} \rangle  ^2 \nonumber
\end{eqnarray}

We define the finite volume susceptibility and their analog for the higher order 
derivatives of the free energy (zero momentum renormalized couplings) 
\begin{equation}
\chi^{(q)}_n\equiv\frac{\langle \left(\sum_{2^n {\text sites}} \phi_x  \right)   ^q\rangle ^c}{2^n}
\end{equation}
We restrict our considerations to the set of initial values such that 
the infinite volume limit of $\chi^{(2l)}_n$ exists and is finite for every 
positive $l$. This means that we are not at another critical point or more generally not on a critical hypersurface 
at the boundary of the HT phase. We emphasize that the existence of the infinite volume limit
requires $0<c<2$. For $c>2$, the 
energy of a constant field configuration scales faster than the number of sites and the model has no interest from a statistical mechanics point 
of view. 

In the following, we assume that the initial values $a_{0,l}$ are such that,
\begin{equation}
\lim_{n\rightarrow \infty }\chi^{(q)}_n=\chi^{(q)}\ ,
\label{eq:lim}
\end{equation}
is finite. From Eq. (\ref{eq:acinter}), it is then clear that 
for $n$ large enough, we have the leading scaling
\begin{equation}
a^c_{n,l}\propto(2 (\frac{c}{4})^l)^n=\lambda_{(l)}^n \ .
\label{eq:aclead}
\end{equation}
It it thus tempting to find a simple relationship between $a^c_{n,l}$ and 
$y_{n,l}$. Indeed, such relation can be found at lowest non-trivial order 
from Eq. (\ref{eq:lowestnt})
which implies that
\begin{equation}
a_{l}^c=y_l+{\mathcal O}(\beta^{l+1})\ .
\label{eq:aclow}
\end{equation}
This can be seen either by using the M\"obius inversion formula
\begin{eqnarray}
y_l &=& \sum_{{\mathbf i}:{\mathcal I}_1({\bf i})=l}
(-1)^{{\mathcal I}_0({\bf i})-1}\times
{({\mathcal I}_0({\bf i})-1)!}
\nonumber \\
& \times & \prod_m \left(
\sum_{{\mathbf r}:{\mathcal I}_1({\mathbf r})=m}
\frac{\prod_jy_j^{r_j}}{r_j!}
   \right)^{i_m}\ \frac{1}{i_m!},
\end{eqnarray}
or more simply by noticing that
\begin{equation}
{\mathnormal e}^{\sum_{l=1}^{\infty}y_lk^{2l}}=\sum_{{\mathbf r}} 
k^{2{\mathcal I}_1({\mathbf r})} \prod_j\frac{y_j^{r_j}}{r_j!}\ .
\end{equation}
Similar formulas are used in multiparticle scattering theory \cite{polyzou80,kowalski81}

Eq. (\ref{eq:aclow}) means that there are no nonlinear contributions of order $\beta ^l$ to $a_l^c$. For instance, there are no $y_1^3$ or $y_1y_2$ terms in $a_3^c$. 
This is expected because the nonlinear terms of order  $\beta ^l$ scale faster than 
$y_l$,  (assuming $0<c<2$). We we say that a term ''scale faster", we mean that it 
goes to zero at a slower rate when $n$ becomes large.
In general, at each RG step, a term $\prod_m y_m^{i_m}$ of order  $\beta ^l$ is multiplied by 
\[2^{{\mathcal I}_0({\mathbf i})}(\frac{c}{4})^l > \lambda_{(l)}=2(\frac{c}{4})^l .\]
The strict inequality comes from the fact that for the nonlinear terms ${\mathcal I}_0({\mathbf i})>1$. 
It is thus clear that nonlinear terms of order $\beta^l$ would spoil the HT
scaling of Eq. (\ref{eq:aclead}) and contradict the existence of a infinite volume limit.

For higher order terms, the sign 
of the denominator $D_{l,{\mathbf i}}$ introduced in Eq. (\ref{eq:deno}) tells us 
whether or not the term scales faster or slower than the linear term. With our sign 
convention,
$c>c_{crit.}({l,{\mathbf i}})$, means $D_{l,{\mathbf i}}<0$ and the term spoils the 
HT scaling Eq. (\ref{eq:aclead}). Since the coefficients are rational functions 
of $c$, they cannot vanish suddenly when $c$ becomes larger than $c_{crit.}({l,{\mathbf i}})$. Consequently if $0<c_{crit.}({l,{\mathbf i}})<2$, the coefficient of the corresponding term 
is expected to vanish identically. 

We have checked that this argument is consistent with our previous explicit calculations. We have used Eqs. (\ref{eq:conna}), (\ref{eq:aofh}) 
and the already calculated coefficients in Eq. (\ref{eq:nlexph}) to calculate
\begin{equation}
a_{l}^c = y_l+\sum_{{\mathbf i}: {\mathcal I}_1({\mathbf i})>l} t_{l,{\mathbf i}} y_{1}^{i_{1}}
y_{2}^{i_{2}} \ldots \ ,
\label{eq:nlexpac}
\end{equation}
up to order 7. For all the 50 terms 
with $0<c_{crit.}<2$, we found that the corresponding $t_{l,{\mathbf i}}$ 
are identically zero. 

\section{The HT phase}
\label{sec:infvol}

In the previous section, we have argued (and checked explicitly up to order 7) 
that terms that scale faster than the 
linear term for $c_{crit.}<c<2$ have a zero coefficient. In this section, 
we discuss more carefully some aspects of the argument 
and explain that having such terms nonzero would result in serious inconsistency.

First of all, the existence of a HT phase is well established. The existence of a
infinite volume limit \cite{miracle} and the absence of spontaneous magnetization for sufficiently high temperature \cite{dyson69} can be shown rigorously for $0<c<2$ and a Ising measure. Bounds on the free energy density \cite{parisi88}, can be established 
for $0<c<2$ and measures with a compact support.
The argument should also apply to measures 
that can be well approximated by measures with a compact support (see Eq. (\ref{eq:vol}) and the argument \cite{convpert} that for Landau-Ginzburg measures, the restriction 
to $|\phi|<\phi_{max}$ leads to exponentially controllable errors). 

It is thus reasonable to assume that there exists some neighborhood of the HT fixed point where 
the infinite volume limit of the susceptibility and higher order derivatives 
(see Eq. (\ref{eq:lim})) exist. 
Terms scaling faster than the linear term seem to contradict the existence of these 
infinite volume limits. However, we should 
exclude the possibility that several terms (scaling identically) 
cancel each others. The existence of universal ratio of amplitudes means that we cannot in general pick arbitrary initial values for the scaling variables.
However, such constraints apply 
for large values of the HT scaling variables. On the other hand, for arbitrarily small 
values of the HT scaling variables, one should be able to make independent variations of each variable
while staying in the HT phase. 
This prevents the fine-tuning required to obtain cancellations.
The HT fixed point $R^{\star}=1$ corresponds to a local measure $W(\phi)\propto
\delta(\phi)$ for which the correlations are zero. 
It is intuitively clear that by taking measures narrowly peaked at zero, one
can avoid long range correlations. This continuity argument can probably be made 
rigorous by using 
Banach spaces as in Refs. \cite{miracle,kw88}.
We conclude that the coefficients $t_{l,{\mathbf i}}$ in Eq. (\ref{eq:nlexpac}) 
of the terms with 
$0<c_{crit.}(l,{\mathbf i})<2$ must vanish identically. 

\section{The absence of poles for $0<c<2$}
\label{sec:inte}

We are now in position to show that the small denominator problem can be evaded for 
any $c$ such that $0<c<2$ and that the solution of the RG flows problem suggested in Eq. (\ref{eq:solve}) can be constructed safely order by order. 
In section \ref{sec:conn}, we have constructed the $a_l^c$ in terms of the 
previously calculated $a_l$. However we could have proceeded directly, writing 
the $a^c_{n+1,l}$ in terms of the $a_{n,l}^c$:
\begin{equation}
a^c_{n+1,l}={\mathcal M}_l^k a^c_{n,k}+\sum_{k+q\geq l} v_l^{kq}a^c_{n,k}a^c_{n,q}+\dots	
\end{equation}
The coefficients $v_l^{kq}$ and the higher order ones can 
be obtained by using the expansion of Eq. (\ref{eq:conngen}) in the logarithm of Eq. (\ref{eq:rec})
and expanding order by order in ${\mathbf a}_n^c$.
The series does not terminate. The linear transformation is the same as before because
$a_l^c$ and $a_l$ only differ by nonlinear terms.
Using \begin{equation}
a_{n,l}^c = {\mathcal R}^r_l h_{n,r}^c\ ,
\label{eq:aofhc}
\end{equation}
we obtain
\begin{equation}
h^c_{n+1,l}=\lambda_{(l)} h^c_{n,l}+ \sum_{k+q\geq l}w_l^{kq}h^c_{n,k}h^c_{n,q}+\dots	
\end{equation}
We then introduce the expansion
\begin{equation}
h_{l}^c = y_l+\sum_{{\mathbf i}: {\mathcal I}_1({\mathbf i})>l}s^c_{l,{\mathbf i}} \prod _m y_m^{i_m} \ ,
\label{eq:hcnlexp}
\end{equation}
and obtain
\begin{equation}
s^c_{l, \mathbf{i}} = \frac{N^c_{l,{\mathbf i}}}{D_{l,{\mathbf i}}}
\label{eq:sc}
\ .
\end{equation}
with $N^c_{l,{\mathbf i}}$ given by a formula similar to Eq. (\ref{eq:num}), except
that it does not terminate. A detailed analysis shows that the 
two formulas have in common that the numerator depends 
only on coefficients of strictly lower orders in $\beta$, and Eq. (\ref{eq:sc}) can be used order by order 
to construct the $s^c_{l, \mathbf{i}}$ for generic values of $c$.

Since ${\mathcal R}^{-1}$ is upper triangular, we see from Eq. (\ref{eq:aofhc}) that $h_l^c$ is equal to $a_l^c$ plus terms which go to zero faster. Consequently, 
for large $n$, the leading scaling is 
\begin{equation}
h^c_{n,l}\propto \lambda_{(l)}^n \ .
\label{eq:hclead}
\end{equation}
Following reasonings used before, this implies that terms 
in the expansion Eq. (\ref{eq:hcnlexp}) 
that scale faster than $y_l$ for any $0<c<2$ should have a vanishing coefficient.
In other words:
\[ 0<c_{crit.}(l,{\mathbf i})<2 \Rightarrow s^c_{l,{\mathbf i}}=0 \ .\]
Given the specific form of the $s^c_{l,{\mathbf i}}$ given in Eq. (\ref{eq:sc}), the $h_l^c$ have no poles for $0<c<2$. The $a_l^c$ being linear combinations of $h_l^c$ and the $a_l$ being linear combinations of products of 
$a_l^c$, we conclude that the expansion of the $a_l$ in terms of the 
scaling variables have also no poles for $0<c<2$, 
in agreement with the conjecture stated in section \ref{sec:hofy}. 

Again we see that there exists a unique continuous definition of the scaling variables that 
can be used at particular values of $c$ where the denominator is exactly zero.
From a practical point of view, the calculation at fixed $c$ of the $s^c_{l, \mathbf{i}}$ is easier
than the calculation of the $s_{l, \mathbf{i}}$, because 
no limit needs to be taken explicitly. The $s^c_{l, \mathbf{i}}$ being rational function of $c$ they cannot be 
zero everywhere except at isolated values. Consequently, we can set to zero the $s^c_{l, \mathbf{i}}$ having 
$c_{crit.}(l,{\mathbf i})<2$ even at values of $c$ where $D_{l,{\mathbf i}}=0$.

\section{The initial value problem}
\label{sec:init}
 
We now return to Eq. (\ref{eq:solve}). In order to complete our solution of the problem, 
namely expressing the ${\mathbf a}_n$ in terms of their initial values
${\mathbf a}_0$, we need to calculate ${\mathbf y}({\mathbf a}_0)$. 

Before doing this, we want to show that the initial values ${\mathbf y}_0$ 
have a very simple interpretation. We have learned in the previous sections that 
$y_{n,l}$ is the \textit{ only} leading term of $a_{n,l}^c$ when $n$ becomes large.
If at a given $0<c<2$, a nonlinear terms scales exactly like $y_{n,l}$, then by 
increasing $c$ slightly (but keeping $c<2$), we can make this term dominant 
in contradiction with the existence of the infinite volume limit.
Consequently,
\begin{equation}
\lim_{n\rightarrow\infty}\lambda_l^{-n}a^c_{n,l}=\lim_{n\rightarrow\infty}\lambda_l^{-n}y_{n,l}=y_{0,l} \ .
\end{equation}
From Eq. (\ref{eq:acinter}), we see that
\begin{equation}
y_{0,l}=(- \beta)^l \frac{1}{2l!}\chi^{(2l)}\ .
\label{eq:yint}
\end{equation}
This means that the infinite volume limit of the susceptibility and of the 
the higher derivatives of the free energy density completely determine the RG flows 
in the HT phase. This also means that the calculation of $y_{0,l}$ given $a_{0,l}$ 
is nontrivial far away from the HT fixed point. However, we can take advantage of the 
fact that 
\begin{equation}
\lambda_{(l)}^{-n}y_{n,l}=y_{0,l} \  
\end{equation}
to estimate $y_{0,l}$ using expansions valid at intermediate values of $n$.

We now discuss the inversion question. We need to determine the coefficients 
$r_{l,{\mathbf i}}$ of the expansion
\begin{equation}
y_{l} = h_l+\sum_{{\mathbf i}} r_{l,{\mathbf i}} \prod _m h_m^{i_m} \ .
\label{eq:nlexpy}
\end{equation}
This can be done by replacing the $y_l$ appearing in the expansion of the $h_l$ 
in Eq. (\ref{eq:nlexph}) by Eq. (\ref{eq:nlexpy}). 
This yields an Eq. of the form
\[ r_{l,{\mathbf i}}+ s_{l,{\mathbf i}}+ X_{l,{\mathbf i}}=0\ ,\]
with $X_{l,{\mathbf i}}$ linear in the ${\mathbf s}$ and multilinear in ${\mathbf r}$
of strictly lower order. One can then construct the $r_{l,{\mathbf i}}$ order by order 
without ever creating a pole in the range $0<c<2$. 
At lowest non-trivial order, we have
\begin{equation}
h_l=\sum_{{\mathbf i}:{\mathcal I}_1({\bf i})=l}(-1)^{{\mathcal I}_0({\bf i})-1}({\mathcal I}_0({\bf i})-1)!\prod_m \frac{y_m^{i_m}}{i_m!}+ {\mathcal O}(\beta^{l+1}) \ .
\end{equation}
A more detailed analysis shows that for higher orders
\begin{equation}
r_{l,\mathbf i}=(\frac{c}{c-4})^{{\mathcal I}_1({\mathbf i})-l}Y_{l,\mathbf i}(c)\ ,
\end{equation}
with $Y_{l,\mathbf i}(c)$ having poles only for $2\leq c<4$. The values of $Y_{l,\mathbf i}(c)$ up to order 4 are given in Table \ref{table:ql2}.
\begin{table*}
\caption{\label{table:ql2}Values of $Y_{l,\mathbf i}(c)$, and $T_{l,\mathbf i}(c)$ defined in the text.}
\begin{ruledtabular}
\begin{tabular}{|c|c|c|c|}
\hline
$l$ & $ \prod _m h_m^{i_m}$ & $Y_{l,\mathbf i}(c)$ & $-T_{l,\mathbf i}(c)$  \\
\hline
\rule{0mm}{1mm}& &  \\
1&${{h_1}}^2$ & $-\left( \frac{2 + c}{-2 + c} \right) $ & $-2 + c$\\
\rule{0mm}{1mm}& & \\
 1&  ${{h_1}}^3$ & $\frac{160 - 224c + 44c^2 + 4c^3 + c^4}
      {{\left( -2 + c \right) }^2\left( -8 + c^2 \right) }$ & $-4 + c^2$\\
      \rule{0mm}{1mm}& & \\
   1& ${h_1}{h_2}$ & $\frac{3\left( -40 + c^2 \right) }{-8 + c^2}$ & $
     -8 + c^2$\\
     \rule{0mm}{1mm}& & \\
    1&  ${{h_1}}^4$ & $
     -\left( \frac{-229376 + 638976c - 364544c^2 - 160768c^3 + 
          109056c^4 + 11648c^5 + 1664c^6 - 5952c^7 + 392c^8 + 
          148c^9 + 6c^{10} + c^{11}}{{\left( -2 + c \right) }^3
          \left( -8 + c^2 \right) \left( -32 + c^3 \right) 
          \left( -16 + c^3 \right) } \right) $ & $-8 + c^3 $\\ 
           \rule{0mm}{1mm}& & \\
        1&  ${{h_1}}^2{h_2}$ & $\frac{-6
        \left( 24576 - 53248c - 19456c^2 + 10496c^3 + 3968c^4 + 
          800c^5 - 576c^6 - 40c^7 + 2c^8 + c^9 \right) }{
        \left( -2 + c \right) \left( -8 + c^2 \right) 
        \left( -32 + c^3 \right) \left( -16 + c^3 \right) }$ & $-16 + c^3$ \\   
         \rule{0mm}{1mm}& & \\  
   1& ${{h_2}}^2$ & $\frac{-1536}{-32 + c^3}$ & $-32 + c^3$\\  \rule{0mm}{1mm}& & \\  
   1& ${h_1}{h_3}$ & $-15$ & $-32 + c^3$ \\ 
   \rule{0mm}{1mm}& & \\ 
\hline
\rule{0mm}{1mm}& & \\
2&${{h_1}}^3$ & $\frac{6 + c}{-2 + c}$ & $-1 + c$\\  
  \rule{0mm}{1mm}& & \\
  2& ${h_1}{h_2}$ & $-\left( \frac{14 + c}{-2 + c} \right) $ & $-2 + c$\\  
   \rule{0mm}{1mm}& & \\
   2&${{h_1}}^4$ & $\frac{-\left( 608 - 1504c + 356c^2 + 36c^3 + 
          3c^4 \right) }{2{\left( -2 + c \right) }^2
        \left( -8 + c^2 \right) }$ & $-2 + c^2$\\ 
        \rule{0mm}{1mm}& & \\
     2&${{h_1}}^2{h_2}$ & $\frac{-2
        \left( -704 + 1216c - 316c^2 - 20c^3 + c^4 \right) }{
        {\left( -2 + c \right) }^2\left( -8 + c^2 \right) }$ & $-4 + c^2$\\ 
        \rule{0mm}{1mm}& & \\
   2&${{h_2}}^2$ & $\frac{3\left( -104 + c^2 \right) }{-8 + c^2}$ & $-8 + c^2$\\ 
   \rule{0mm}{1mm}& & \\
   2&${h_1}{h_3}$ & $\frac{-240}{-8 + c^2}$ & $-8 + c^2$\\  
   \rule{0mm}{1mm}& & \\
\hline
\rule{0mm}{1mm}& & \\
3&${{h_1}}^4$ & $-\left( \frac{10 + c}{-2 + c} \right) $ & $
     -\left( \frac{1}{2} \right)  + c$ \\ \rule{0mm}{1mm}& & \\
   3&${{h_1}}^2{h_2}$ & $\frac{2\left( 18 + c \right) }{-2 + c}$ & $-1 + c$\\ \rule{0mm}{1mm}& & \\
    3& ${{h_2}}^2$ & $\frac{-16}{-2 + c}$ & $-2 + c $\\ \rule{0mm}{1mm}& & \\
   3& ${h_1}{h_3}$ & $-\left( \frac{22 + c}{-2 + c} \right) $ & $-2 + c$\\
     \rule{0mm}{1mm}& & \\
\rule{0mm}{1mm}& &  \\
\end{tabular}
\end{ruledtabular}
\end{table*}  

\section{The HT expansion}
\label{sec:ht}

A simple application of the method presented here is the calculation of the high-temperature 
expansion at finite volume. As a simple example, we consider the first order
correction to the susceptibility for a Ising measure ($R_0(k)=\cos(\sqrt{\beta}k)$). Using the results found in the previous sections, we obtain
\begin{eqnarray}
\chi_n^{(2)}=&\frac{-2}{\beta}& a_{n,1}(\frac{c}{4})^n\nonumber \\
=& \frac{-2}{\beta}& \left[ y _{0,1}+\frac{2c}{2-c}(\frac{c}{2})^n \  (y_{0,1})^2 \right. \nonumber \\
&\  & \ \left. +\frac{6c}{4-c}(\frac{c}{4})^n\  y_{0,2} \right]+{\mathcal O}(\beta^2)\ \nonumber\\
\equiv &1& + \beta b_{1,n}+{\mathcal O}(\beta^2)\ .
\end{eqnarray}
Using $a_{0,1}=\frac{-\beta}{2}$ and $a_{0,2}=\frac{\beta^2}{24}$, we obtain 
\begin{eqnarray}
y_{0,1}&=&-\frac{\beta}{2}-\frac{\beta^2 c}{4(4-c)}
-\frac{\beta^2 c(2+c)}{4(4-c)(2-c)}\nonumber \\
y_{0,2}&=&-\frac{\beta^2 }{12}\nonumber \ ,
\end{eqnarray}
and consequently
\begin{eqnarray}
b_{1,n}=&\ &
\frac{2c}{(4-c)(2-c)}-\frac{c}{2-c}(\frac{c}{2})^n \nonumber \\ &+&\frac{c}{4-c}(\frac{c}{4})^n \ .
\end{eqnarray}
This is in agreement with results obtained \cite{meuricejmp95} 
using graphical methods.

\section{RG invariants}
\label{sec:rginv}

In Hamiltonian mechanics, integrable systems with $q$ degrees of freedom have $q$ constants of motions and $q$ periodic 
variables with independent periods depending on the constants of motion. In the present case, time is discrete and 
exponential decays replace the quasiperiodic behavior. For a truncation of dimension $l_{max}$, it is nevertheless
possible to construct $l_{max}-1$ constants of motion:
\begin{equation}
G_l\equiv- (2l)!\frac{y_{n,l}}{(-2y_{n,1})^{(l-1)(D/2)+l}}
\end{equation}
These quantities are $n$-independent and we call them RG invariants.
We can calculate them at $n=0$. Using Eq. (\ref{eq:yint}), we obtain
\begin{equation}
G_l=(-1)^{l+1}\frac{\beta ^{\frac{D}{2} (1-l)}\chi^{(2l)}}{(\chi^{(2)})^{(l-1)(D/2)+l}}	
\end{equation}
We can also calculate them at large enough values of $n$ where the HT expansion works well.
The minus sign has been introduced in order to have $G_l>0$ for $D=3$.

We now concentrate on the unstable direction of Wilson's fixed point. 
We set the relevant scaling variable associated with this fixed point to a value $u$ which becomes 
our coordinate along the unstable direction and we set all the 
irrelevant ones to zero.  We call $G_l(u)$ the corresponding value of the ratio.
Given that $u$ is a scaling variable and that $G_l$ is RG invariant, we have 
\begin{equation}
G_l(\lambda _w u)=G_l(u)\ ,
\end{equation}
with $\lambda_w $ the eigenvalue corresponding to the unstable direction of Wilson's 
fixed point. Consequently, we have the Fourier expansion:
\begin{equation}
	G_l(u)=\sum_rA_{l,r}u^{i r \omega} \ ,
\end{equation}
with
\begin{equation}
\omega =\frac{2\pi}{\ln \lambda	_w}\ .
\end{equation}
If the oscillatory terms are  
very small, as noticed in Refs. \cite{osc1,osc2,dual}, we have the approximate
universal ratios 
\begin{equation}
G_l(u)\simeq A_{l,0}\ .
\end{equation}
These constants can be calculated in an intermediate region where the expansions 
in both scaling variables are 
valid \cite{jsp02}.

\section{Conclusions}

We have shown that the scaling variables corresponding to the HT fixed point of Dyson's hierarchical 
model can be constructed order by order without small denominator problems. The ambiguity noticed before \cite{smalld} for calculations at fixed values of $c$ can be raised by requiring the continuity in $c$. 
Practical calculations at finite $c$ are most easily done by following the
explicit construction sketched in section \ref{sec:inte} for the connected part 
where no complicated limit is required. The remaining poles for $2\leq c\leq 4$ reflect the degeneracy of the
linear spectrum at $c=4$ or the fact that some finite size corrections become leading effects 
for some value of $2\leq c<4$ (where the infinite volume limit does not exist).

We have solved the linear problem in compact form but at this point no compact form 
is available for the nonlinear problem. Even though we have ``constants of motion" 
(the RG invariants), we do not have simple expressions for them (as the integrals 
of integrable models). The question of the convergence of the series remains to be 
addressed and should result in the construction of the boundary of the HT phase. 
Finally, it would be desirable to extend the method to models with $\eta \neq 0$.
One way to achieve this goal would be to 
develop methods to systematically improve the hierarchical approximation.

\begin{acknowledgments}
This research was supported in part by the Department of Energy
under Contract No. FG02-91ER40664 and also by a 
Faculty Scholar Award at The University of Iowa and a residential
appointment at the Obermann Center for Advanced Studies at the
University of Iowa. 
We thank the participants of the joint Math-Physics seminar at the 
University of Iowa for interesting suggestions.
We thank the
Aspen Center for Physics for its hospitality while
this work was initiated.
\end{acknowledgments}


\end{document}